\documentstyle[11pt, epsfig]{article}
\newcommand{\blankline}{\vskip .3cm}
\newcommand{\f}{\begin{equation}}
\newcommand{\ff}{\end{equation}}
\newcommand{\bea}{\begin{eqnarray}}
\newcommand{\eea}{\end{eqnarray}}

\begin{document}
\centerline{\LARGE  Nonperturbative dynamics}
\blankline
\centerline{\LARGE  for abstract (p,q) string networks}
\rm
\blankline
\blankline
\centerline{Fotini Markopoulou  
and Lee Smolin}
\blankline
\blankline
\blankline

\centerline{ \it  Center for Gravitational Physics and 
Geometry}
\centerline{\it Department of Physics}
 
\centerline {\it The Pennsylvania State University}
\centerline{\it University Park, PA, USA 16802}
 \vfill
\centerline{December 2, 1997}
\vfill
\centerline{ABSTRACT}
We describe abstract (p,q) string networks which
are the string networks of Sen without the information about
their embedding in a background spacetime.  The non-perturbative
dynamical formulation invented for {\it spin} networks,  in terms of
causal evolution of dual triangulations, is applied on them. 
The formal transition amplitudes are
sums over discrete causal histories
that evolve (p,q) string networks.  The dynamics
depend on two free SL(2,Z) invariant functions which describe the
amplitudes for the local evolution moves.   
 
\blankline
email addresses: fotini@phys.psu.edu, smolin@phys.psu.edu 
\eject
\section{Introduction}

In a very interesting paper \cite{sennet},  Sen introduces the
notion of a network of $(p,q)$ strings and hypothesizes that
they might play a role in a non-perturbative formulation of
string theory analogous to the role $SU(2)$ spin networks
play in non-perturbative quantum general 
relativity\cite{sn1,volume1}.  These are composed of
(p,q) strings that meet at trivalent vertices introduced by
Schwarz\cite{schwarz} and shown to be BPS states
in \cite{3bps}.  This suggestion has been
explored in several recent papers\cite{kl,morestringnets}.  
In this note we take up Sen's suggestion and
show that the  (p,q) string networks fit naturally into
a class of generalizations of spin networks that 
have been studied as possible non-perturbative quantum theories
of gravity\cite{algebraic}.  
The dynamics of spin networks that one of us suggested in
\cite{fotini1} may be applied directly to (p,q) string networks
to yield a formal non-perturbative background independent 
formulation  of their dynamics.  This formulation is in turn
closely related to  path integral formulations
of non-perturbative quantum gravity based on deformations of
topological quantum field theories\cite{mike,carlomike,louis,
louisjohn,baez,barbieri}.  

Penrose originally  invented spin networks to provide a simple 
realization of the idea that the quantum geometry of
spacetime must
be discrete and based only on algebra and combinatorics, with
all continuum notions arising in the continuum limit\cite{roger-sn}.
As introduced by Sen in \cite{sennet}, (p,q) string networks depend
on an embedding in a flat background $9+1$ dimensional spacetime.
However if they are to serve as the basis of a non-perturbative
formulation of string theory the (p,q) string networks should be
removed from this context and postulated to exist as background
independent combinatorial entities from which the continuum
description is to be derived as a suitable approximation.  
The possibility of this is suggested by the striking fact that,
just as in Penrose's original formulation \cite{roger-sn}, angles
for the embedding of the (p,q) string 
network in flat space are derived
purely from the combinatorics of the network\cite{sennet}.
  
This raises the question of how the dynamics and observable
algebra of the theory are to be defined in the absence of a 
background manifold.  
In \cite{fmls1,fotini1,algebraic} we have studied this problem for
non-embedded extensions of spin-networks.  This led to the
introduction of a general theory with two basic features:
(1)  The spin networks are extended to 2-dimensional labeled
surfaces and the space of states is constructed algebraically.
(2)  Dynamics is expressed in terms of local moves that
generate histories as combinatorial structures which 
share several of the
properties of Lorentzian spacetimes including causal structure
and many-fingered time.

As shown by Schwarz \cite{schwarz} and elaborated
for networks 
by Krogh and Lee \cite{kl},  (p,q) strings
can be understood as labeled 2-dimensional surfaces.  
This leads directly to the application of the dynamical formulation
introduced in \cite{algebraic,fotini1} to string networks.

We must emphasize that we have not yet shown 
whether the dynamics we
propose has any connection with the dynamics of (p,q)
string networks embedded in background manifolds.   The dynamics
in the form proposed here is specified by two free $SL(2,Z)$
invariant functions.  In the future, these may be determined by the
condition that they match the dynamics predicted by $M$
theory, but this has not yet been done.   

In the next section we introduce the
abstract (p,q) string networks, which obey  the combinatorics
of the (p,q) networks of Sen \cite{sennet} but are not
embedded in any background manifold.    
In section 3 we
describe  non-perturbative dynamics for these abstract
(p,q) networks.

\section{Abstract (p,q) string networks}

Spin networks (and their extensions to general quantum
groups) can be understood in the general framework
first developed for conformal field 
theory
and topological field theory \cite{segal,atiyah,category}
in which states associated to
various manifolds are constructed from
algebraic operations.  
It is easy to see that (p,q) string networks can be formulated
in the same framework.

We consider sets $\{(p_i,q_i)\}$ consisting of $n$ relatively prime 
pairs of integers.  
Each pair may be visualized as the first homology 
classes of a $T^2$ as in \cite{kl},
but the only role this will play in the abstract formalism
is that the torus parameter $\tau$ may come into the
$SL(2,Z)$ invariant functions that specify the dynamics.  
We may note that the relatively prime pairs (p,q) also
label the rational numbers or, alternatively, the 
projective representations of the maps of the 
circle to itself\footnote{
We thank Louis Crane for this observation.}.  
As a result, there is a natural multiplication defined on them,
which is 
\f
(p_1,q_1) \otimes (p_2,q_2) = (p_1+p_2,q_1+q_2) = r 
(p_3,q_3) \rightarrow (p_3,q_3 )
\label{prod}
\ff
where $r$, a positive integer, is the greatest common factor
of $p_1+p_2$ and $q_1+q_2$.  Associativity follows from the
uniqueness of prime factorization.  

\begin{figure}
\centerline{\mbox{\epsfig{file=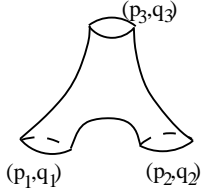}}}
\caption{The representation of the product (1) as a 
three string vertex.} 
\label{pqtrinions}
\end{figure}

The basic idea of a categorical construction of a $TQFT$ or
conformal field theory\cite{segal,atiyah,category} is that
sets of labels, which are representations of some algebra,
are mapped to manifolds such that the product is represented
by cobordisms.  In this case we will associate the sets
$\{(p_i,q_i)\}$ to sets of circles 
so that the algebra they generate 
is mapped to cobordisms of two manifolds.
The product (\ref{prod}) is symbolized
by a trinion, or 3-punctured sphere, with an object
$(p,q)$ associated with each incoming puncture and
$(-p,-q)$ associated with the one outgoing puncture  as in Figure 
(\ref{pqtrinions}).   When $r\neq 0$ we say there exists
a morphism from $(p_1,q_1) \otimes (p_2,q_2) $
to $(p_3,q_3)$.    
The product (\ref{prod}) is thus the algebraic counterpart of the
three string vertex, which when the strings are represented
as tubes becomes the trinion.  
Under reversal of orientation of a puncture we require
$(p,q) \rightarrow (-p,-q)$.  Taking all the orientations consistently
a trinion is labeled by three relatively prime 
pairs $(p_i,q_i)$ such that,
\f
\sum_{i=1}^3 p_i = \sum_{i=1}^3 q_i =0
\label{good}
\ff
A trinion with labels $(p_i,q_i)$ satisfying (\ref{good}) will be
called a good trinion.  

\begin{figure}
\centerline{\mbox{\epsfig{file=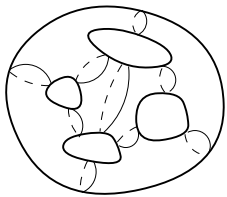}}}
\caption{A genus 4 surface cut into six trinions $B^3_I$ by circles
$c_\alpha$.   } 
\label{stringnet}
\end{figure}

An abstract $(p,q)$ string network is then
described by a compact surface
$\cal S$ with $m$ punctures constructed from sewing 
 good trinions together along the punctures,
as shown in Figure (\ref{stringnet}).  The result is a surface $\cal S$ with
a set of circles $c_\alpha$  which
decompose it into a set of good trinions, with labels $(p,q)_\alpha$
on the punctures $c_\alpha$ of the trinions.  
A closed $(p,q)$ string network is one without free punctures.

A non-degenerate string network will be one that has a trinion
decomposition in which no two trinions meet on more than one
circle. We may note that the string networks of Sen \cite{sennet}
are non-degenerate because the strings are
straight lines in flat spacetime.  We will thus restrict ourselves
to the consideration of non-degenerate abstract string networks.

To the set of non-degenerate 
abstract $(p,q)$ string networks we associate a
Hilbert space ${\cal H}^{SN}$ constructed as follows.  Given each
surface $\cal S$ with its circles
$c_\alpha$ there is a space ${\cal V}^{\cal S}_{c_\alpha}$
spanned by a basis of states 
$| {\cal S},c_\alpha; (p,q)_\alpha \rangle$.  We choose the
inner product $\langle | \rangle $ so that these are orthonormal
\cite{algebraic}.
We then define
\f
{\cal H}^{SN} = \bigoplus {\cal V}^{\cal S}_{c_\alpha}
\ff
 where the sum is over all labeled surfaces 
$({\cal S},c_\alpha)$.  
\footnote{ 
All of this can be expressed compactly by saying that there
is a functor from the cobordism category
$Cob$ whose objects are
sets of circles and whose morphisms are two manifolds with
boundary to the category of relatively
prime pairs of integers with the product (\ref{prod}).
The general formulation of this kind of correspondence
is described in \cite{atiyah,segal,category}.}

What distinguishes theories of the type we are about to propose
from 3-d topological quantum field theory is that there are 
operators which act on ${\cal H}^{SN}$ to change the topology
of the surface $\cal S$.  
 
For example, consider $\Upsilon^1$ and $\Upsilon^2$ to be two 
open $(p,q)$ string
nets, each with $n$, free punctures, with
labels $(p,q)_k$, $k=1,...,p$.  Thus  $\Upsilon^1$ and $\Upsilon^2$
have the same boundary, but are otherwise different.
There is then an operator ${\cal C}_{\Upsilon^1, \Upsilon^2}$
which acts on each state in the basis 
$| {\cal S},c_\alpha;(p,q)_\alpha\rangle$
by looking for components of $({\cal S},(p,q)_\alpha)$ which
are isomorphic to $({\Upsilon}^1,(p,q)_k)$, and
replaces it by a different 2-surface $(\Upsilon^2,(p,q)_k)$, 
which has {\it the same boundary}, 
but otherwise differs.

\begin{figure}
\centerline{\mbox{\epsfig{file=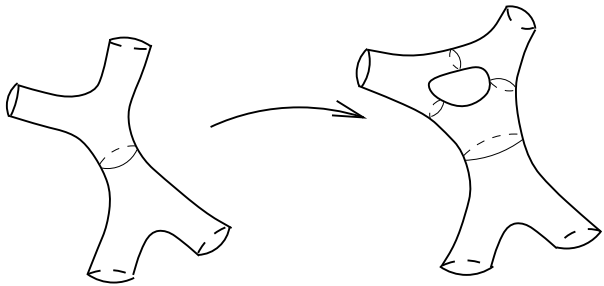}}}
\caption{A local substitution move.} 
\label{pqmove}
\end{figure}

Now, there may be more than one places in $({\cal S},(p,q)_\alpha)$
where $({\Upsilon}^1,(p,q)_k)$ is recognized as a subset. There are then 
several maps 
\f
r_I :({\Upsilon}^1,(p,q)_k) \rightarrow ({\cal S},(p,q)_\alpha) .
\ff
For each $I$ the map $r_I$ picks out a set of
$n$ non-intersecting circles $c_k^I$, $k=1,...,n$ in
$\cal S$.  Cutting $\cal S$ on these circles decomposes it into
the two pieces $r_I ({\Upsilon^1})$ and $\left({\cal S}- 
r_I ({\Upsilon}^1)\right)$.  We may then sew $\Upsilon^2$ onto
$\left({\cal S}- 
r_I ({\Upsilon}^1)\right)$ along the circles $c_k^I$ as they 
are also identical with the set of punctures of $\Upsilon^2$.  
We may call this state 
$|\left ( {\cal S}-r_I(\Upsilon^1)\right )  \cup \Upsilon^2\rangle$
with the dependence on the circles and labelings understood.
The operator ${\cal C}_{\Upsilon^1, \Upsilon^2}$ is then defined
by 
\f
{\cal C}_{\Upsilon^1, \Upsilon^2} |{\cal S} \rangle =
\sum_I |\left ( {\cal S}-r_I(\Upsilon^1) \right ) \cup \Upsilon^2\rangle
\ff
A slightly more complicated definition can be made when
$\cal S$ is not compact.  An example of such a substitution move
is given in Fig. (\ref{pqmove}).
These operators are local as they only change
local pieces of a string network which are isomorphic to a given
piece.  We now show that a particular set of such
operators may be used to formulate the dynamics of the
abstract $(p,q)$ string networks.

\section{Dynamics}

Following the proposal made for causal evolution of spin networks
in \cite{fotini1} we may give a general form for 
dynamics of (p,q) string networks that is locally causal.  We note
that in a background independent formalism the notion of locality
must be based on the states themselves.  

\begin{figure}
\centerline{\mbox{\epsfig{file=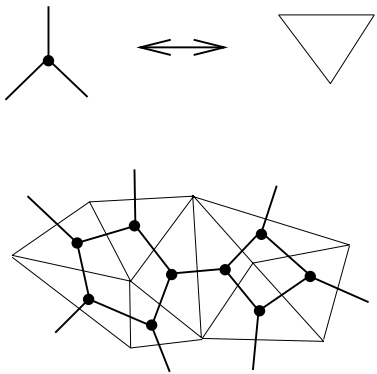}}}
\caption{Triangulations dual to trivalent networks} 
\label{2+1dual}
\end{figure}

For the $2+1$ case it is actually more convenient to describe
the dynamics in the dual picture \cite{fotini1}.
An abstract trivalent string network $|{\cal S},j_\alpha \rangle $ 
may be equivalently described  by a labeled
triangulation  ${\cal T}$, (See Fig. (\ref{2+1dual}))
in which the sides of the triangles carry the
$(p,q)$ labels and  the condition (\ref{good}) is satisfied by
the three $(p,q)$ pairs around each triangle.  
If the string network is closed, the corresponding triangulated surface
is compact. We may also note that non-degeneracy of the string network
implies that the triangulation is non-degenerate in the sense that
no two triangles share more than one edge.  
The issue of the evolution of a network is now translated to the 
evolution of a 2-dimensional triangulation. Now, the generating {\it local}
evolution moves from a non-degenerate 
2-dimensional triangulation to another
one, which leave invariant the topology of the triangulated surface, 
are the
Pachner moves (\cite{pachner}, which were defined 
originally for $PL$ manifolds).
These have already been used to describe causal spacetime histories of
spin networks \cite{fotini1} (and in the euclidean case they can 
be found in
\cite{carlomike,baez}. Let us use them here to write 
down a formal path-integral 
evolution for abstract $(p,q)$ networks. 
 
\begin{figure}
\centerline{\mbox{\epsfig{file=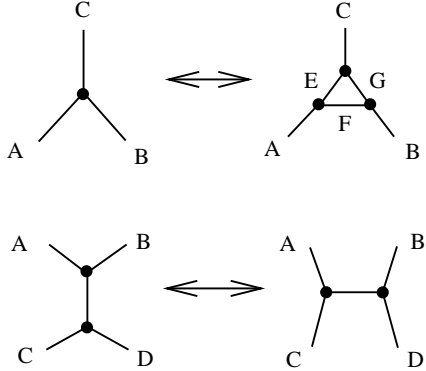}}}
\caption{The $2+1$ dimensional Pachner moves.} 
\label{2+1pachner}
\end{figure}

The Pachner moves are transitions from an initial $(p,q)$ triangulation
${\cal T}_i$ to a final one ${\cal T}_f$. They are labelled tetrahedra 
having the triangles of ${\cal T}_i$ and ${\cal T}_f$ as their boundaries. 
The different moves are the different ways of gluing a tetrahedron on
$n$ connected triangles of ${\cal T}_i$ and get $4-n$ connected triangles in
${\cal T}_f$, $n\leq 3$. The new edges which are created
by the move must be labeled consistently with (\ref{prod})
given the labelings of the remaining edges. See Fig.(\ref{2+1pachner}). 
For $(p,q)$ networks, these moves are particularly simple. 
The $1\rightarrow
3$ move is always possible.  The inverse, $3\rightarrow 1 $ requires that 
there are three triangles in ${\cal T}_i$ around a single vertex,
as in Fig. (\ref{2+1pachner}).  
A $2\rightarrow 2$ move is only possible when labels of the four external 
sides 
add up to $(0,0)$.  

By iterating the moves one constructs a sequence of string networks
$\Gamma_0, \Gamma_1, ...= \{ \Gamma_i \}$.  
As in general relativity in which local
evolution of initial data generates a spacetime, any such
sequence generates a {\it history}. Also as in general relativity, a
given history may be generated many ways by local evolution moves.
For this reason it is best to define it abstractly, as follows.

A history $\cal M$ between
an initial string network $\Gamma_0$ and a final string network 
$\Gamma_f$ is then a three dimensional simplicial complex
consisting of $N$ tetrahedra $\tau_I$ such that 
$\partial {\cal M}= \Gamma_0 \cup \Gamma_f$ (where we 
identify the string net with its dual triangulation). Each edge of
$\cal M$ is labeled by a pair $(p,q)$ such that the conditions
(\ref{good}) are satisfied on every face.  The faces of each history
make up a causal set\footnote{A causal set is a partially ordered set,
where the order relation stands for causality, 
which has no closed causal loops.
The formulation of discrete quantum theories
of gravity with such causal structures has been studied
by Sorkin and collaborators\cite{rafael} and 'tHooft\cite{th}.
} that is defined as follows.  The faces of
each tetrahedra are divided into a past set $\cal P$ and a 
future set, $\cal F$
so that for two faces $f_1 \in {\cal P}$ and $f_2\in {\cal F}$ of
the same tetrahedron 
we have $f_1 < f_2$ (where $<$ indicates the partial ordering).  
We then extend this to all faces in
$\cal M$ so that $f_1 < f_r$ if there is a sequence of $r$ faces
$f_i$ such that $f_i < f_{i+1}$ within some tetrahedron.  
Thus each history has a discrete causal 
structure.

Given the causal structure of a history $\cal M$  
there is a natural definition of a spacelike surface.
A discrete spacelike surface, $\Delta$,  is 
a collection of triangles in ${\cal M}$ without boundary
and with  no two triangles causally related. 
Because the causal structure is local, each 
history $\cal M$ can be foliated in many different ways
into sequences of spacelike surfaces (abstract $(p,q)$ 
networks),  ${\cal T}_i,\Delta_1,...,\Delta_I,...,{\cal T}_f$.
If a history was generated by a sequence $\Gamma_i$ of string
networks each of these is a spacelike slice, but the history
$\cal M$ they construct will have many spacelike slices not
in the list $\{ \Gamma_i \}$.  Thus, the theory has a discrete
analogue of the multi-fingered time of general relativity.
 
To each tetrahedron $T$ in ${\cal M}$ we can associate
an amplitude, which is the amplitude for a local transition
between two string networks.  
The amplitude ${\cal A}_{T_I}[\tau,(p_i,q_i)_I]$ 
will then depend on the labels of the edges of
that tetrahedron.  To be consistent with string theory it should
be some $SL(2,Z)$ invariant function of the 
the string theory parameter
$\tau$ and the $(p,q)$ integers on 
its faces. There will also be in general different amplitudes
for different causal structures on the tetrahedron, $T_I$, this
is indicated by the dependence on $T_I$ in the form of the
amplitude ${\cal A}_{T_I}[\tau,(p_i,q_i)_I]$.

\begin{figure}
\centerline{\mbox{\epsfig{file=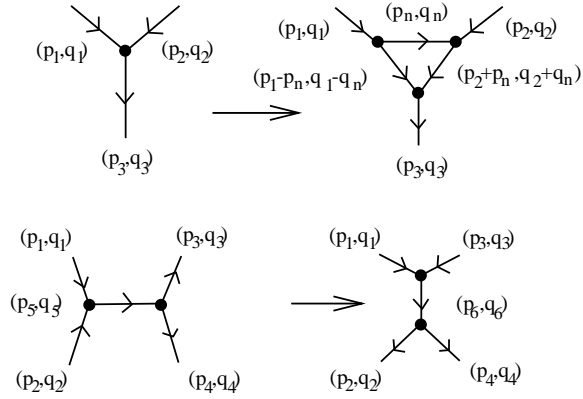}}}
\caption{The allowed Pachner moves for $(p,q)$ string networks.} 
\label{allowed}
\end{figure}

If the history ${\cal M}$ contains $N$ tetrahedra $T_I$, 
the amplitude for $\cal M$ is local when it is the product
\f
{\cal A}[{\cal M}] =\prod_{I=1}^N {\cal A}_{T_I}[\tau,(p_i,q_i)_I] .
\ff
For the $1\rightarrow 3$ Pachner move, 
${\cal A}_{1\rightarrow 3}[\tau , p_i,q_i ]$ 
must be an 
$SL(2,Z)$ invariant function of the parameter $\tau$,
the three 
incoming labels $(p_i,q_i)$, $i=1,2,3$, symmetric in the
$i$, and an internal pair $(p_n,q_n)$ that runs around
the internal loop, subject
to (\ref{good}). (See Fig. (\ref{allowed}).)  The amplitude 
for the inverse 
$3\rightarrow 1$ moves is by hermiticity the complex conjugate
of the $1\rightarrow 3$ moves.  
For the $2\rightarrow 2$ move the amplitude 
${\cal A}_{2\rightarrow 2}[\tau , p_i,q_i ]$
is a real  
function of $\tau$ and the four external edges
$(p_i,q_i)$, subject to the constraint that
$\sum_{i=1}^4 (p_i,q_i) =(0,0)$. (See Fig. (\ref{allowed}).) 
An invariant   choice of 
${\cal A}_{1\rightarrow 3}[\tau , p_i,q_i ]$ 
and $ {\cal A}_{2\rightarrow 2}[\tau ,p_i, q_i ]$ 
defines a non-perturbative $(p,q)$ network theory.
Using the operators of the last section, a hermitian
operator $\widehat{H}$ on ${\cal H}^{SN}$ can be defined \cite{algebraic}
such that $<{\cal T}^2|\widehat{H} |{\cal T}^1 > $ vanishes unless
${\cal T}^2$ differs from ${\cal T}^1$ by the action of one
Pachner move, in which case it is equal to the corresponding
amplitude, ${\cal A}_{T_I}[\tau,(p_i,q_i)_I]$.

The result of the action $\widehat{H} |{\cal T}_i\rangle$ is then 
the linear
combination of the states reached by the 
possible Pachner moves from $|{\cal T}_i\rangle$
each multiplied by the corresponding amplitude  
${\cal A}_{T_I}[\tau,(p_i,q_i)_I]$
\footnote{For the details of this construction
in the general case, see
\cite{algebraic}.}.
Given ${\cal T}_i$ and ${\cal T}_f$ then, the transition 
amplitude between the two is 
\f
{\cal A}_{{\cal T}_i\rightarrow {\cal T}_f}
= \langle {\cal T}_i| e^{\imath t H }| {\cal T}_f\rangle.
\label{evolve}
\ff
Expansion of 
(\ref{evolve}) gives the amplitude as a sum over histories
$\cal M$.  $t$ is a formal parameter which scales the
amplitudes and has no interpretation as a physical time.
(One measure of  physical time is the number, $N$, of 
tetrahedra in each history, as this is related, in the
continuum limit, to an integral of a 
positive function over the spacetime history.)  We note that
formally the evolution defined by (\ref{evolve}) is unitary.  

We do not know anything more about the choices of the
amplitudes.  It is possible that consistency with perturbative
string theory plus $SL(2,Z)$ invariance will constrain or
determine them\footnote{We thank Shyamoli Chaudhuri and
Djordje Minic for conversations about this.}.
Alternately, one may try to search the space of amplitudes
for choices that lead to critical behavior, resulting in a continuum
limit.  The appropriate critical behavior should be related to
directed percolation, for reasons explained in \cite{fmls1}.
There is also a result that suggests that perturbative string theories
may be derived from the perturbation theory of theories of the
kind we have considered here, in the case that they have a good continuum 
limit\cite{stringsfrom}.

\section{Conclusion}

What we have described here is a proposal for a formal path integral 
evolution of abstract $(p,q)$ networks
based on two steps.  The first is the construction of a space of
states ${\cal H}^{SN}$ composed of abstract graphs with the
same labels and combinatorics as a large class
of $BPS$ states which are networks of (p,q) strings
embedded in $9+1$ dimensional Minkowski space.    
This fits into the categorical framework behind much recent
work in non-perturbative quantum gravity and topological
field theory.   The second step is a proposal for the dynamics
of the theory based on a set of local moves.
The exact dynamics of the theory is not prescribed, but its
choice reduces to the specification 
of two $SL(2,Z)$ invariant functions.

\section*{ACKNOWLEDGEMENTS}
We are grateful to Shyamoli Chaudhuri,  
and Djordje Minic for conversations about this work
and its possible connection to string theory, to Louis Crane
for helpful information and to Sameer Gupta and Ashoke Sen 
for pointing out a silly mistake in the first draft of this paper.
This work was supported by
NSF grant PHY-9514240 to The Pennsylvania State
University and a NASA grant to the Santa Fe Institute.


\begin{thebibliography}{99}

\bibitem{sennet}A. Sen {\it String networks},
hep-th/9711130.

\bibitem{sn1}C. Rovelli and L. Smolin,  
``Spin networks and quantum gravity"  
gr-qc/9505006, Physical Review D 52 (1995) 5743-5759.

\bibitem{volume1}C. Rovelli and L. Smolin
{\it Discreteness of area and volume in quantum gravity}
 Nuclear Physics B 442 (1995) 593.  Erratum: Nucl. Phys.
B 456 (1995) 734.

\bibitem{schwarz}J. H. Schwarz, {\it Lectures on Superstrings
and M theory dualities} hep-th/9607201, Nucl. Phys.
Proc. Suppl. 55B (1997) 1.

\bibitem{3bps}K. Dasgupta and S. Mukhi, {\it BPS nature of
3-string junction} hep-th/9711094.

\bibitem{kl}M. Krogh and S. Lee, {\it String networks
from M theory} hep-th/9712050.

\bibitem{morestringnets}S.-J. Rey and J.-T. Yee {\it BPS
dynamics of triple (p,q) string junction}
hep-th/9711202

\bibitem{algebraic}F. Markopoulou and L. Smolin
{\it Quantum geometry with intrinsic local causality}
preprint, Dec. 1997, gr-qc/9712067.

\bibitem{fotini1}F. Markopoulou,  
{\it Dual formulation of spin network evolution},
 gr-qc/9704013, CGPG preprint (1997).
 

\bibitem{mike}M. Reisenberger,
{\it A lattice worldsheet sum for 4-d Euclidean general 
relativity} , gr-qc/9711052.

\bibitem{carlomike}M. Reisenberger and C. Rovelli,
{\it ``Sum over Surfaces'' form of Loop Quantum Gravity},
gr-qc/9612035

\bibitem{louis}L. Crane, {\it A Proposal for the Quantum 
Theory of Gravity} gr-qc/9704057. 

\bibitem{louisjohn}J. Barrett and L. Crane, {\it
Relativistic spin networks and quantum gravity},
gr-qc/9709028 ; L. Crane, {\it  On the interpretation 
of relativistic spin networks and the balanced state sum},
 gr-qc/9710108.

\bibitem{baez}J. Baez, {\it Spin foam models},
 gr-qc/9709052.

\bibitem{barbieri}A. Barbieri {\it Quantum tetrahedra
and simplicial spin networks} gr-qc/9707010.

\bibitem{roger-sn}R Penrose: in {\it Quantum theory and 
beyond}  ed T Bastin, Cambridge U Press 1971;
in {\it Advances in Twistor Theory}, ed. L. P. Hughston and R. S. 
Ward,
(Pitman,1979) p. 301; in {\it Combinatorial Mathematics and
its Application} (ed. D. J. A. Welsh) (Academic Press,1971);
{\it Theory of quantized directions}
unpublished manuscript.

\bibitem{fmls1}F. Markopoulou and L. Smolin
{\it Causal evolution of spin networks}  gr-qc/9702025.
CGPG preprint (1997), Nuclear Physics B, in press.

\bibitem{stringsfrom}L. Smolin, {\it Strings from perturbations
of causally evolving spin networks} preprint, Dec. 1997.

\bibitem{segal}G. Segal,  {\it Conformal field theory}
oxford preprint (1988).

\bibitem{atiyah}M. Atiyah, {\it Topological quantum field theory}
Publ. Math. IHES 68 (1989) 175; {\it The Geometry and
Physics of Knots}, Lezion Lincee
(Cambridge University Press, Cambridge,1990).

\bibitem{category}J. Baez,  {\it 
An Introduction to n-Categories} ,  q-alg/9705009

\bibitem{pachner}V. Pachner, Europ. J. Combinatorics, 12 (1991) 129.

\bibitem{rafael}L. Bombelli, J. Lee, D. Meyer and
R. D. Sorkin,  {\it Spacetime as a causal set} Phys. Rev. Lett. 
59 (1987) 521.

\bibitem{th}G. 't Hooft,  {\it Quantum gravity: a fundamental 
problem and some
radical ideas.}  Carg\`{e}se Summer School Lectures 1978. 
Publ. ``Recent
Developments in Gravitation''.  Carg\`{e}se 1978. 
Ed. by M. L\'{e}vy
and S. Deser.  Plenum, New York/London, 323;
{\it Quantization of Space and Time in 3 and in 4
Space-time Dimensions}, Lectures held at the NATO Advanced Study
Institute on ``Quantum  Fields and Quantum Space Time", Carg\`ese, July
22 -- August 3, 1996.  gr-qc/9608037; {\it 
The scattering matrix approach for the quantum black
hole: an overview.}  J. Mod. Phys. \underline{A11} (1996) pp.
4623-4688. gr-qc/9607022.
 
\end{thebibliography}
\end{document}